\documentclass{PoS}
\usepackage{lineno}

\newsavebox{\tempbox}

\title{The Kaon identification system at the NA62 experiment at CERN}

\ShortTitle{The NA62 Kaon identification system}

\author{\speaker{Karim Massri}\thanks{On behalf of the NA62 Collaboration: G.~Aglieri Rinella, R.~Aliberti, F.~Ambrosino, R.~Ammendola, B.~Angelucci, A.~Antonelli, G.~Anzivino, R.~Arcidiacono, I.~Azhinenko, S.~Balev
% ...
}\\
        University of Liverpool\\
        E-mail: \email{karim.massri@cern.ch}}

\abstract{The NA62 experiment at CERN SPS aims to measure the branching ratio of the ultra-rare kaon decay $\kpinn{+}$ with 10\% precision, collecting $\sim 100$ events, assuming the Standard Model~(SM) branching ratio of $8.4 \times 10^{-11}$, starting in 2016.
The NA62 experiment uses a kaon decay-in-flight technique and is exposed to a 750~MHz high-energy unseparated charged hadron beam, in which kaons are a minor component (6\%). Kaon identification is therefore mandatory to reduce the interference of the dominant non-kaon component with the experimental measurements. The NA62 kaon identification system and its performances are presented.}

\FullConference{38th International Conference on High Energy Physics\\
		3-10 August 2016\\
		Chicago, USA}

\newcommand{\kpinn}[1]{K^{#1} \to \pi^{#1} \nu \bar{\nu}}

\begin{document}
\section{Introduction}
The NA62 experiment at CERN SPS~\cite{td10} aims to measure the branching ratio of the ultra-rare kaon decay $\kpinn{+}$ with 10\% precision, collecting $\sim 100$ events, assuming the Standard Model~(SM) branching ratio, starting in 2016.
The $\kpinn{+}$ decay is a Flavour Changing Neutral Current~(FCNC) process and therefore it is forbidden in the SM at the tree level.
Furthermore, it is highly CKM-suppressed: $\mathcal{B}(\kpinn{+}) \propto \lambda^{10}$.
These properties make this decay very sensitive to new physics.
The SM expectation is $\mathcal{B}(\kpinn{+}) = (8.4\pm 1.0) \times 10^{-11}$~\cite{bu15}
and the current experimental measurement is $\mathcal{B}(\kpinn{+}) = 1.73^{+1.15}_{-1.05} \times 10^{-10}$~\cite{ar09}, which is based on 7 candidates observed by the E787 and E949 experiments at the Brookhaven National Laboratory~(BNL).
Unlike the experiments~E787 and E949 which used kaons at rest, the NA62 experiment uses high-momentum (75~GeV/$c$) $K^+$ decaying in flight. An advantage of this choice is the higher energy of the decay products, which increases the rejection power for the main background, the $K^+\to\pi^+\pi^0$ decay. A disadvantage of a high-momentum beam is that pions and protons, which dominate the secondary beam, cannot be efficiently separated from kaons.
As a consequence, a particle flux $\sim 17$~times greater than the kaon one passes through the NA62 detector, possibly interfering with the experimental measurements. Positive identification of kaons can reduce the non-kaon contribution to the $\kpinn{+}$ background to an acceptable level,
and is therefore essential to meet the experimental goal.

\section{The NA62 Kaon identification system}
The kaon identification is based on a ChErenkov Detector with Achromatic Ring focus~(CEDAR) placed in the incoming beam. 
CEDAR counters~\cite{bo82} have been constructed and used at CERN since the early '80s for SPS secondary beam diagnostics.
They typically operate in charged-particle beams of MHz flux and have timing resolution of order a few ns.
The NA62 experimental strategy requires the kaon identification system to satisfy the following conditions:
\begin{itemize}
\item[-] kaon tagging efficiency of at least $95\%$;
\item[-] kaon crossing time measurement with a resolution better than 100~ps;
\item[-] pion rejection below $10^{-4}$;
\item[-] capability to perform at a nominal instantaneous rate of 45~MHz (kaons);
\item[-] capability to stand a radiation level of 1 Gy/year.
\end{itemize}
To cope with the challenging 45~MHz kaon rate and to achieve the required time resolution, a new photon detector, called Kaon TAGger~(KTAG), has been built and assembled for the CEDAR.

\subsection{The CEDAR detector}
\label{sec:cedar_optsys}
\begin{figure}[htb]
\begin{center}
\includegraphics[width=0.59\textwidth]{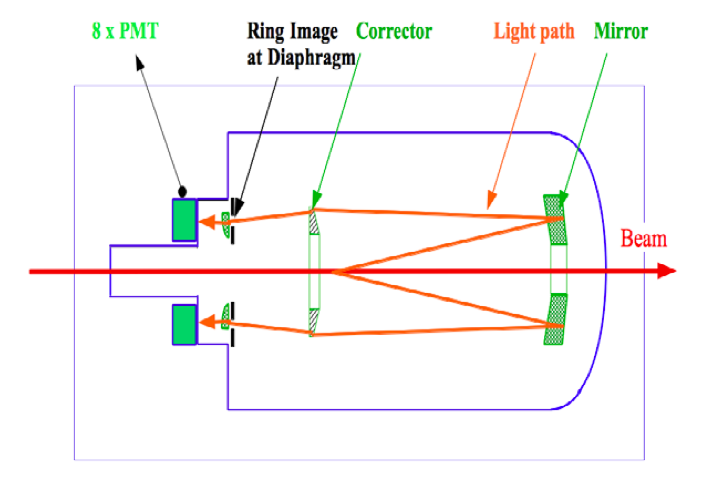}
\includegraphics[width=0.40\textwidth]{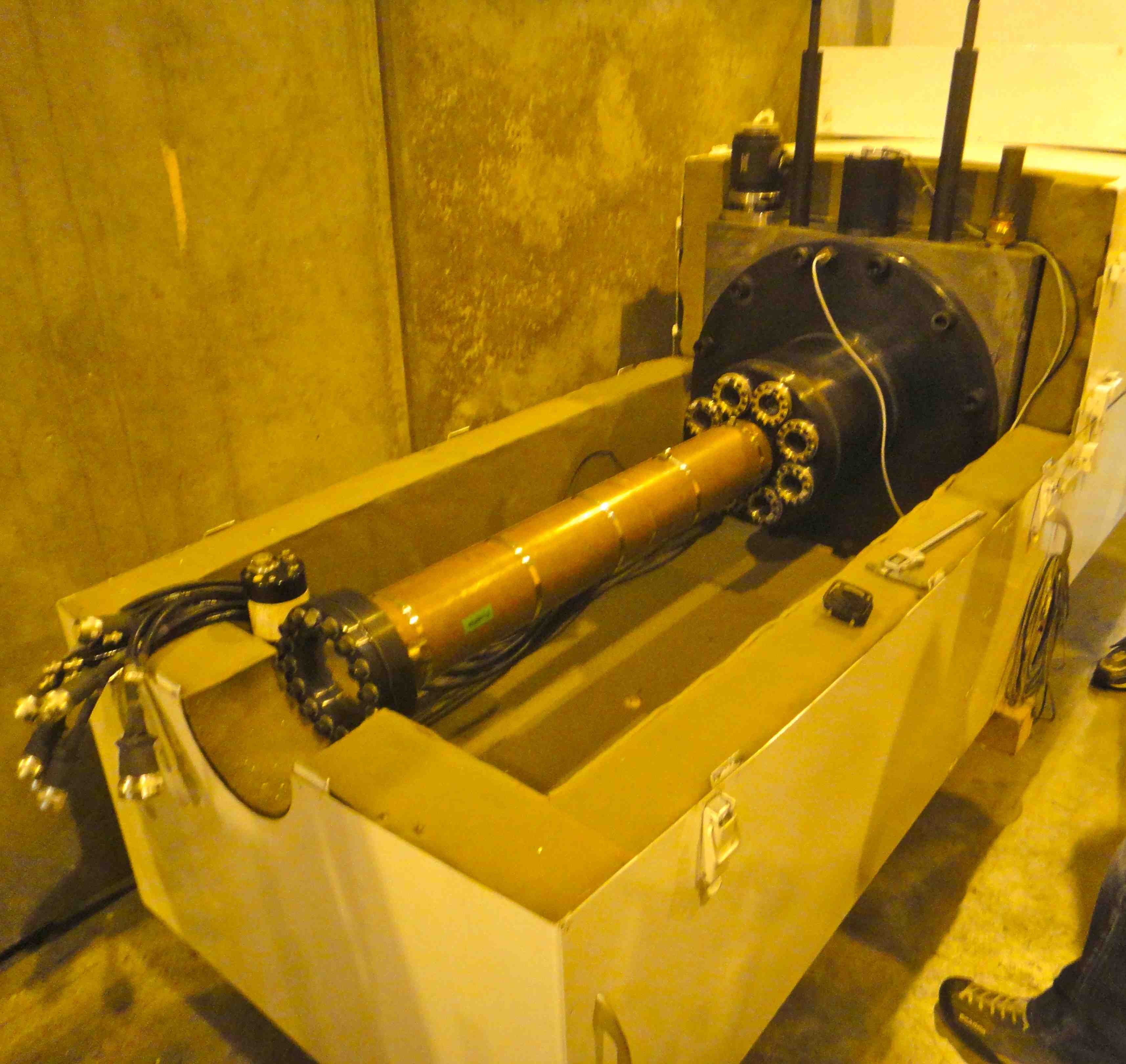}
\caption{a) sketch of the optical system located inside the vessel; b) picture of the upstream end of the CEDAR vessel with the nose and the quartz windows.} \label{fig:CEDARlayout}
\end{center}
\end{figure}

The CEDAR detector is a steel vessel of 55.8~cm external (53.4~cm internal) diameter and 4.5~m length filled with $N_2$ gas and containing an optical system, which is sketched in Fig.~\ref{fig:CEDARlayout}a.
The Cherenkov light emitted by beam particles traversing the vessel is reflected back by a Mangin mirror,
passes through a chromatic corrector lens and arrives to a diaphragm, which selects only light rings within a specified radial range.
The diaphragm has a radius of 100~mm and an adjustable width (from 0 to 20~mm) set to 1.5~mm, optimal for $K/\pi$ separation.
The light passing through the diaphragm arrives to eight quartz spherical windows, which are attached at one end of the vessel and surround the 
nose~(Fig.~\ref{fig:CEDARlayout}b), and reach eight EMI~9820qb Photo-Multipliers~(PMTs), mounted at the upstream end of the quartz windows.
However, the original CEDAR PMTs cannot stand the nominal NA62 kaon rate of 45~MHz: considering the limits on the current drawn from these PMTs it follows that the maximum affordable positively-identified beam particle rate is $\approx 8$~MHz. In addition, their individual time resolution is $\sim2$~ns. Therefore, the kaon time resolution cannot be better than 700~ps, well away from the 100~ps required from the NA62 experimental programme.

\subsection{KTAG: the new photon detector}
\label{sec:cedar_upgrade}
\begin{figure}[h!]
\begin{center}
\includegraphics[width=0.495\textwidth]{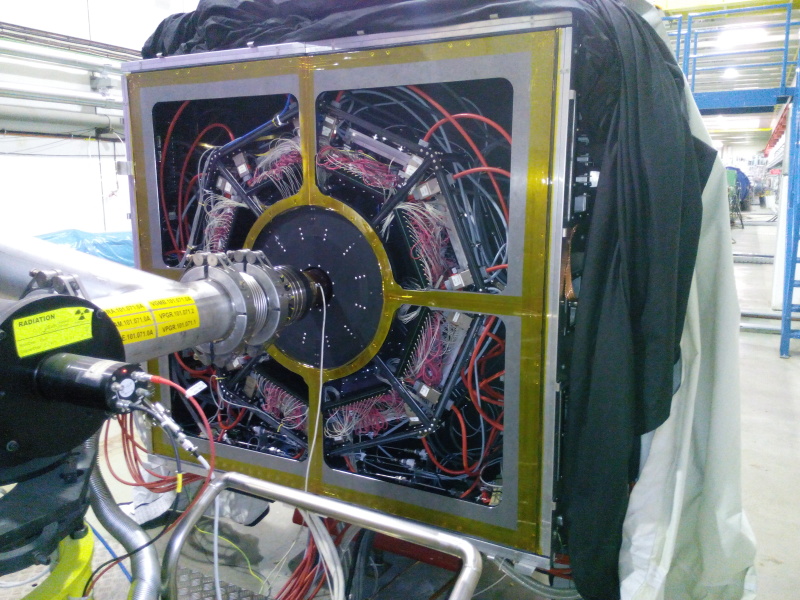}
\includegraphics[width=0.495\textwidth]{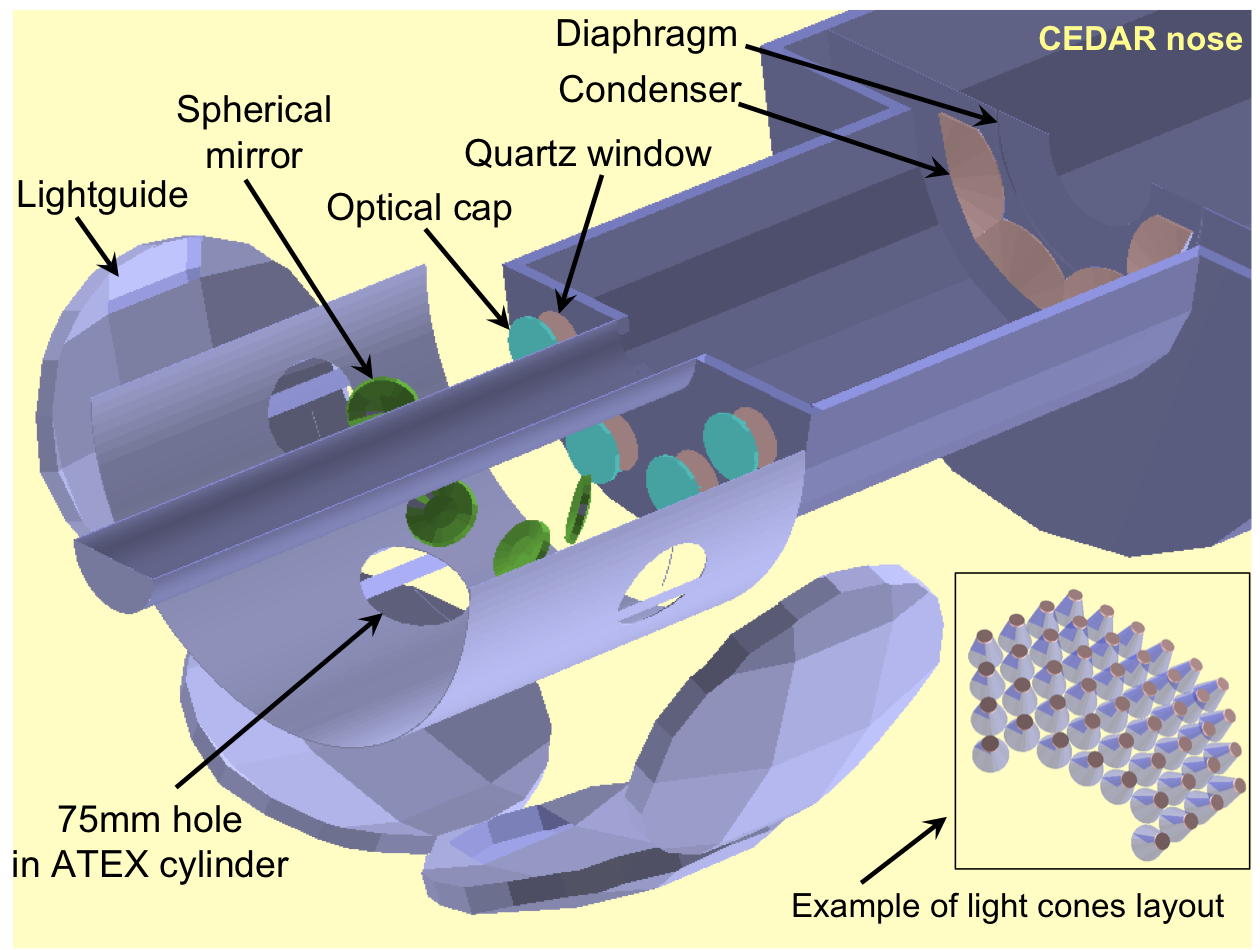}
\end{center}
\caption{a) picture of the new KTAG detector positioned upstream the quartz windows, 
surrounding the nose and housing mirrors, light collection cones and PMTs; b) Screenshot taken from a Geant4 simulation of the KTAG. The components of the optical system outside the CEDAR vessel are sketched.} 
\label{fig:ktag}
\end{figure}

To achieve the NA62 rate and timing requirements, each original CEDAR PMT has been replaced with a high-granularity configuration of \mbox{single-photon} counting photomultipliers.
The new KTAG photon detector~\cite{go15} is shown in Fig.~\ref{fig:ktag}.
An additional structure, spider-shaped with eight arms, is mounted at the upstream end of the vessel, surrounding the nose and the eight quartz windows.
The structure, contained in an additional part referred to as enclosure, is divided into two mirror-symmetric halves and supports the new PMT matrices.
In the following, the space between two arms and the corresponding attached equipment will be referred to as a sector.
Each sector contains a focussing lens (``optical cap'') mounted at one quartz window, a spherical mirror and a solid metallic container (``light box''), hosting light collection cones machined in a solid aluminium plate (``light guide'') and the front-end electronics.
The light collection cones have been designed in order to maximise the light collection efficiency and reduce the dead areas.
The new PMT matrices are equipped with 48 single-anode PMTs: this configuration allows a reduction of the photon rate and average anode current per PMT down to a sustainable level at the nominal NA62 beam intensity. 
The photo-detector technology required for the KTAG must be able to perform single photon counting.
The PMTs chosen for the KTAG design are based on metal package photomultipliers of the Hamamatsu R7400U and R9880U series, 03 and 210 types respectively. 
Both models were selected for their size (8~mm diameter active area) and compactness (16~mm x 12~mm cylindrical shape), time performances on single photon detection ($\sim 300$~ps), resistance to radiation, low dark count rate ($\sim$~few~Hz), sensitivity to visible and near-UV wavelengths.
Each KTAG PMT matrix is equipped with 16 R7400U-03 and 32 R9880U-210 Hamamatsu PMTs.
However, to allow the possibility of future upgrades, the light guide layout has been designed to contain 64 PMTs.

\sbox{\tempbox}{\includegraphics[width=0.45\textwidth, height=0.5\textwidth]{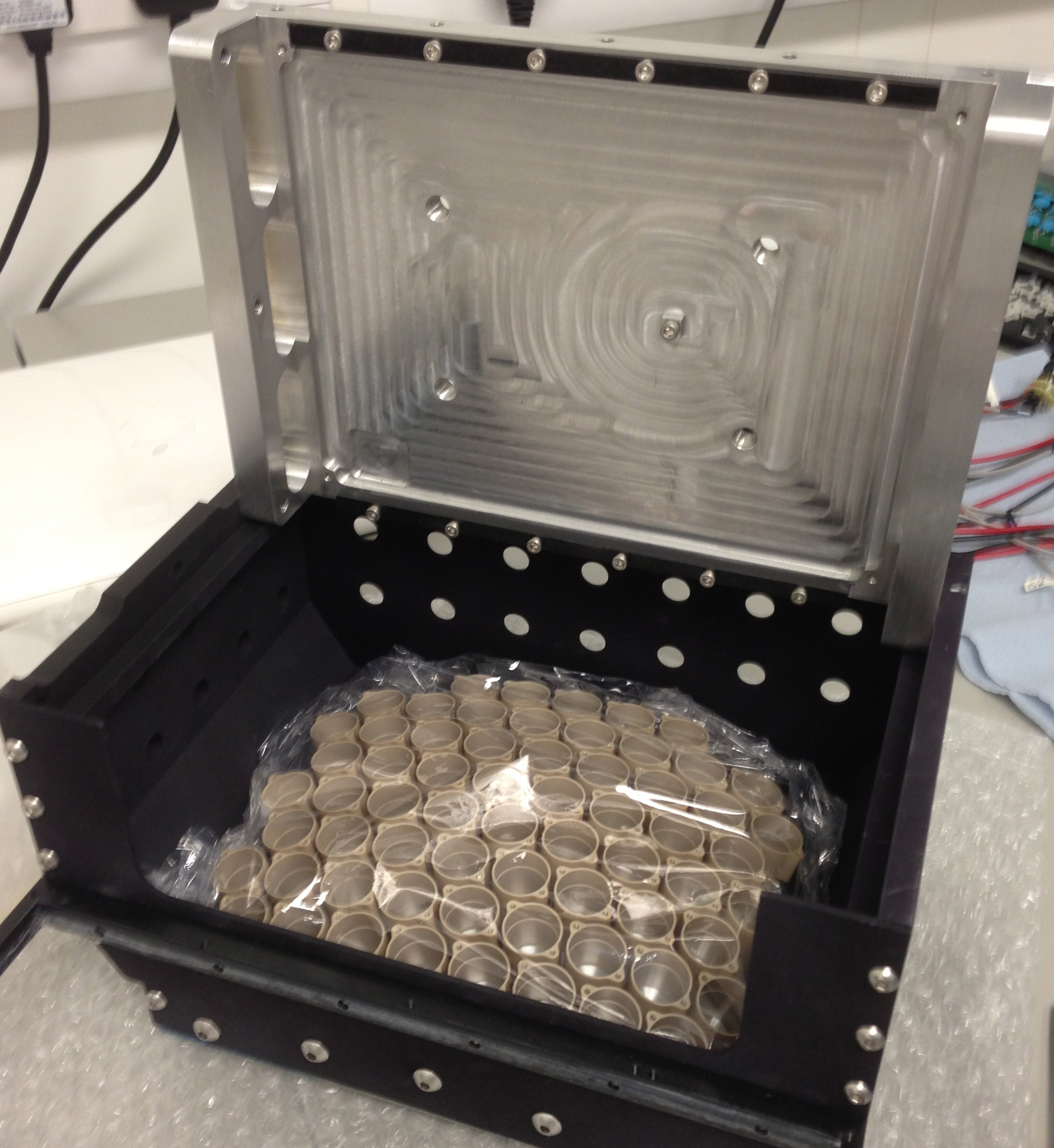}}
\begin{figure}[h!]
\begin{center}
\usebox{\tempbox}\hspace{0.01\textwidth}
\vbox to \ht\tempbox{\vfil \hbox{\includegraphics[width=0.53\textwidth,height=0.4\textwidth]{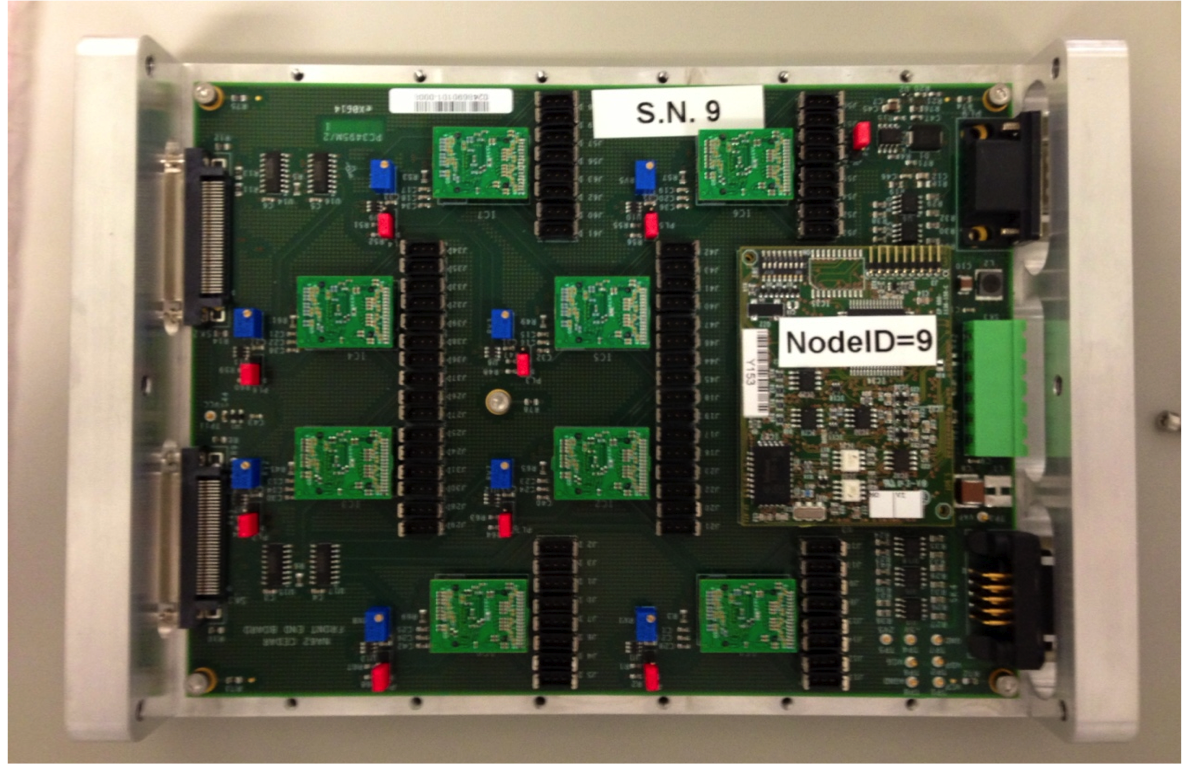}}\vfil}
\end{center}
\caption{a) light box without PMTs, with the metallic lid lifted open and PMT holders attached at the basement. 
b) NINO board with two CAN bus connectors and a central voltage connector mounted at the right side of the board and two LVDS connectors at the left side of the board; the NINO board is screwed to the metallic lid of the light box.} \label{fig:nino-lb}
\end{figure}
The front-end used to interface the PMTs with the readout system relies on NINO ASIC~\cite{an04} (8~channels) mezzanine chips.
In each light box~(Fig.~\ref{fig:nino-lb}a) a board housing 8 NINO mezzanines~(Fig.~\ref{fig:nino-lb}b) is used to shape and discriminate the output signal from PMTs. 
The NINO chip operates in time-over-threshold mode and its threshold can be set remotely and independently for the 8 NINO mezzanine boards.
The NINO digital output is finally sent to the TEL62 motherboards~\cite{an14}, a major upgrade of a FPGA-based readout board inherited from LHCb readout system.

\section{Performances of the NA62 kaon identification system}
\subsection{Kaon time resolution}
\begin{figure}[h]
\begin{minipage}{0.5\textwidth}
\includegraphics[width=\textwidth]{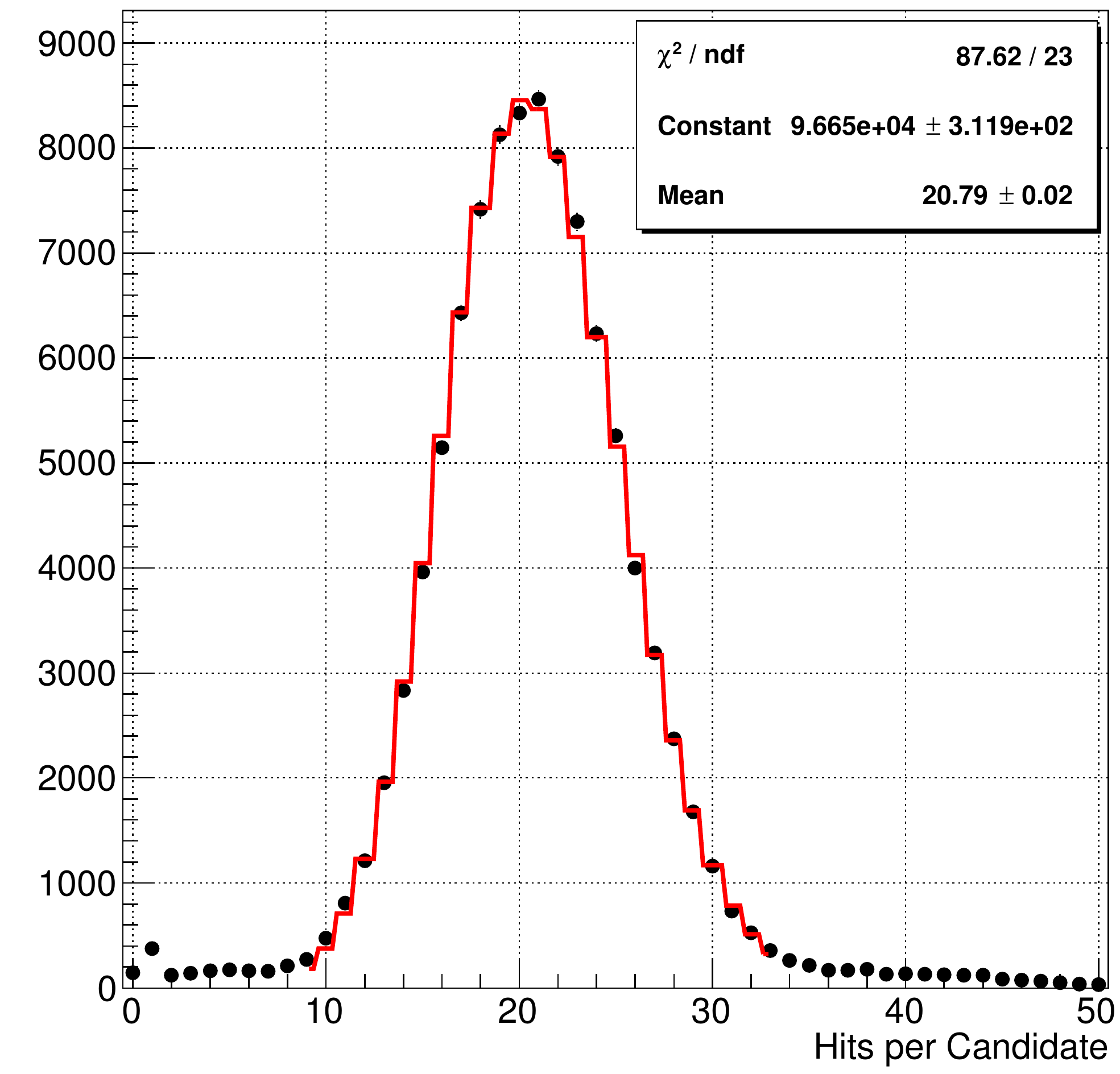}%
\end{minipage}
\hfill
\begin{minipage}{0.5\textwidth}
\includegraphics[width=\textwidth]{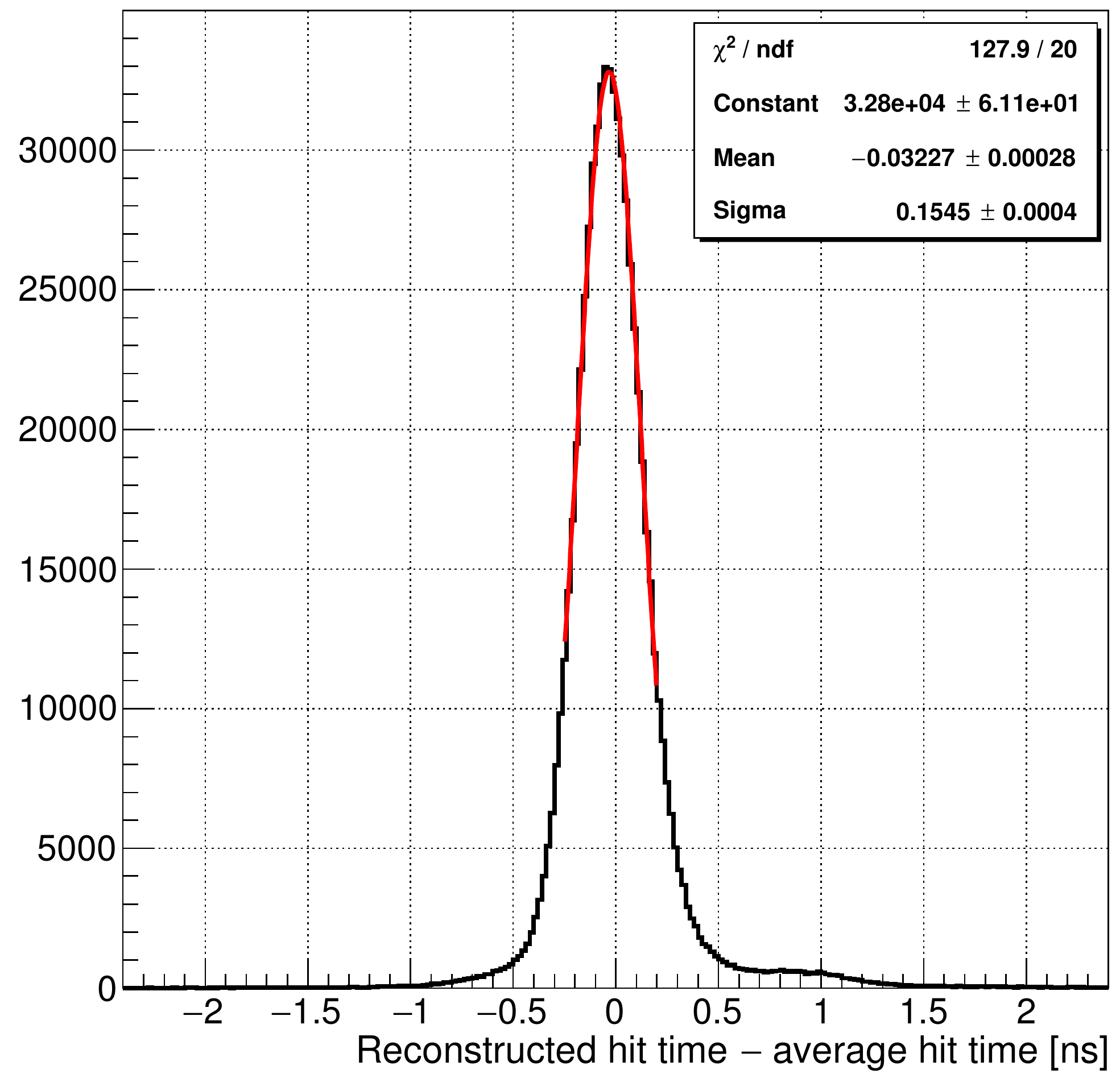}%
\end{minipage}
\caption{a) number of hits per reconstructed kaon, fitted with a Poisson distribution; b) individual PMT time with respect to the kaon~time, fitted with a gaussian distribution.}\label{fig:timeresolution}  
\end{figure}

The CEDAR/KTAG detector has enough degrees of freedom to estimate its own time resolution without relying on an external time reference: the number of hits per kaon (Fig.~\ref{fig:timeresolution}a) is large enough to evaluate the time of the event by performing an average. The residuals with respect to the average~(Fig.~\ref{fig:timeresolution}b) are a measurement of the time resolution of an individual PMT, while the global time resolution can be estimated by dividing by the square root of the number of hits. It is a slightly biased estimate, because of the correlation between average and residuals, but for a number of hits of the order of 10 or more the bias is negligible. The single PMT time resolution, after applying time offsets and time slewing corrections, is about 300~ps (RMS). Since the mean number of hits per kaon candidate is about 20, the kaon time resolution is about 70~ps.
\subsection{$K/\pi$ separation and kaon identification efficiency}
\begin{figure}[h]
\begin{minipage}{0.5\textwidth}
\includegraphics[width=\textwidth]{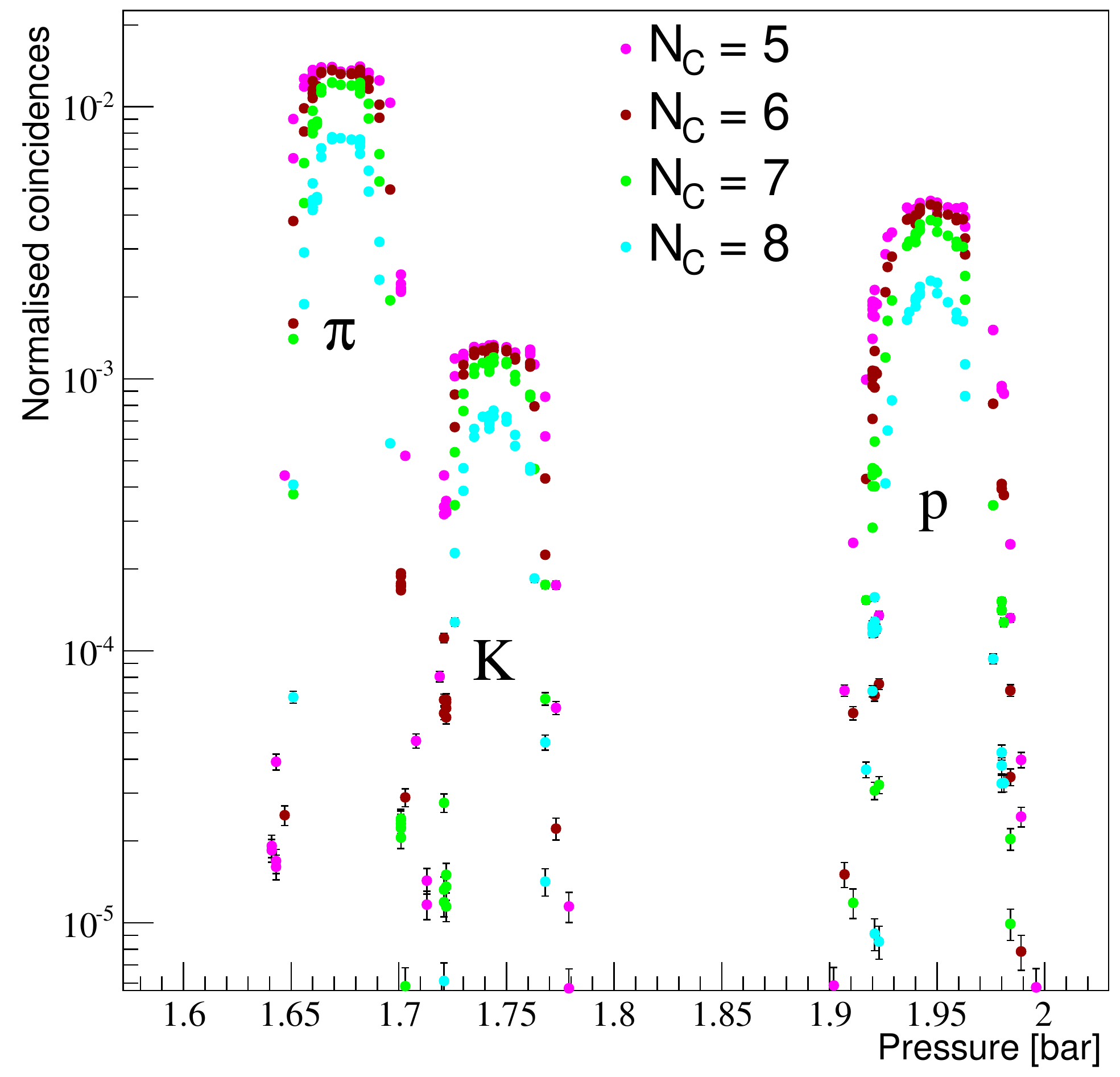}%
\end{minipage}
\hfill
\begin{minipage}{0.5\textwidth}
\includegraphics[width=\textwidth]{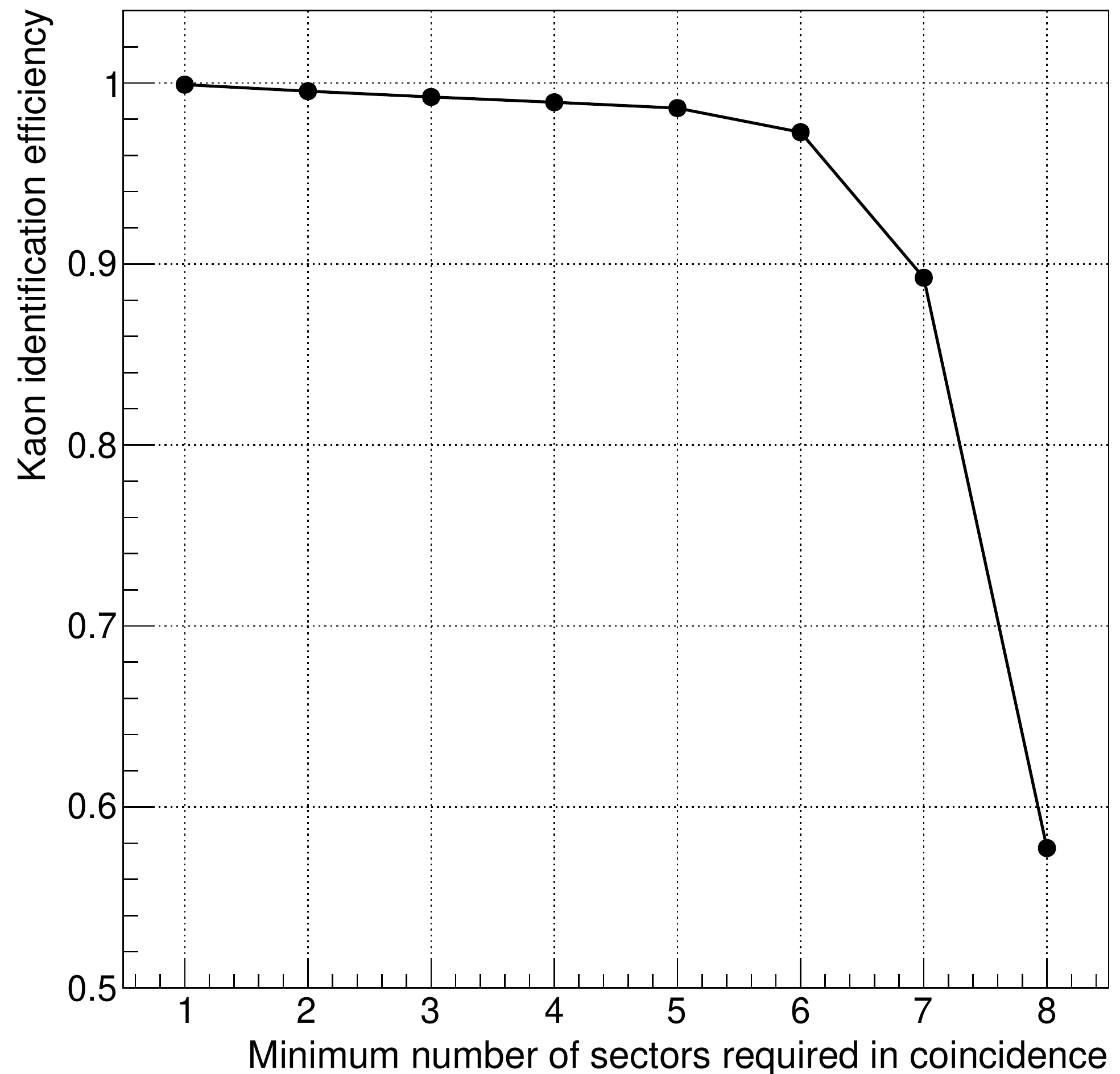}%
\end{minipage}
\caption{a) CEDAR/KTAG response as a function of the $N_2$ radiator pressure; b) Kaon identification efficiency for different requirements on the number of sectors in coincidence.}\label{fig:kaonID}  
\end{figure}

The discriminatory power of the CEDAR/KTAG detector for identifying $\pi^+$, $K^+$ and protons is shown in Fig.~\ref{fig:kaonID}a in terms of the
number of coincidences of signals in the sectors, as a function of the $N_2$ gas pressure.
Three peaks are resolved. Each of them corresponds to a different beam component:
the peak at lower pressure to pions, the intermediate one to kaons and the one at higher pressure to protons. The peak heights reflect the beam composition. Clear separation of the $\pi^+$, $K^+$ and proton peaks is observed by requiring a coincidence of signals in at least 5 sectors.

The kaon identification efficiency is evaluated from a sample of $K^{\pm}\to\pi^+\pi^0$ decays reconstructed without applying any kaon identification requirement. The efficiency measured for different requirements on the number of sectors in coincidence is shown in Fig.~\ref{fig:kaonID}b.

\section{Conclusions}
The NA62 kaon identification system consists of a differential Cherenkov detector upgraded with a new photon detector. The kaon time resolution is measured to be about 70~ps. The kaon identification efficiency measured with reconstructed $K^+\to\pi^+\pi^0$ decays is greater than 98\% when requiring Cherenkov light in coincidence in at least 5 sectors. For the same coincidence requirement, the probability of misidentifying a pion as a kaon while operating at the kaon pressure is estimated to be~$\mathcal{O}(10^{-4})$. 
All the measured performances meet or exceed the NA62 requirements.


\begin{thebibliography}{99}
\bibitem{td10} The NA62 Collaboration, ``NA62: Technical design document'', NA62-10-07 (2010), https://cds.cern.ch/record/1404985.
\bibitem{bu15} A. J. Buras, D. Buttazzo, J. Girrbach-Noe and R. Knegjens, JHEP {\bf 1511} (2015) 033.
\bibitem{ar09} A. V. Artamonov {\it et al.} (BNL-E949 Collaboration), Phys. Rev. {\bf D79} (2009) 092004.
\bibitem{bo82} C. Bovet, R. Maleyran, L. Piemontese, A. Placci, and M. Placidi, CERN-82-13 (1982).
\bibitem{go15} E. Goudzovski {\it et al.}, Nucl. Instrum. Meth. {\bf A801} (2015) 86.
\bibitem{an04} F. Anghinolfi {\it et al.}, Nucl. Instrum. Meth. {\bf A533} (2004) 183.
\bibitem{an14} B. Angelucci {\it et al.}, JINST {\bf 9} (2014) C01055.
\end{thebibliography}
\end{document}